\documentclass[11pt]{article}                 
\usepackage{appb2,epsfig}

\newcommand{\be}{\begin{equation}}
\newcommand{\ee}{\end{equation}}
\newcommand{\bea}{\begin{eqnarray}}
\newcommand{\eea}{\end{eqnarray}}
\newcommand{\bm}{\boldmath}
\newcommand{\ubm}{\unboldmath}
\newcommand{\smallz}{{\scriptscriptstyle Z}} %  a smaller Z
\newcommand{\smallw}{{\scriptscriptstyle W}} %

\newcommand{\fr}{\frac}

\newcommand{\mz}{M_\smallz}
\newcommand{\mw}{M_\smallw}

\def \gev  {\mbox{ GeV}}
\def \mev  {\mbox{ MeV}}

\def \ms   {\overline{\mbox{MS}}}

\def \psl  {p \kern-.45em{/}}
\def \qsl  {q \kern-.45em{/}}
\def \qslov {\overline{q \kern-.45em{/}}}
\def \pslov {\overline{p \kern-.45em{/}}}

\def \sov  {\overline{s}}
\def \mov  {\overline{m}}

\def \Aov  {\bar{A}}
\def \lsim {\raisebox{-.7ex}{$\stackrel{\textstyle <}{\sim}\,$}}

\def \xiw   {\xi_\smallw}
\def \xiz   {\xi_\smallz}
\def \xig  {\xi_\gamma}

\def \LM   {\ln \! \left(\frac{M^2-s}{M^2} \right)}
\def \LMxiw {\ln \!\left(\frac{M^2\xiw-s}{M^2\xiw} \right)}
\def \LS   {\ln \!\left(\frac{\sov-s}{\sov} \right)}
\def \Agamma {A^{\gamma}}

\def \dlr {\Delta r}

\begin{document}
%%%%%%%%%%%%%%%%%%%%%%%%%%%%%%%%%%%%%%%%%%%%%%%%%%%%%%%%%%%%%%%%%%%%%%%%%%%%%%
\eqsec
\bm \title{Radiative Corrections to $W$ and Quark \\
           Propagators in the Resonance Region\thanks{Presented
	   at Zeuthen Workshop on Elementary Particle Theory: Loops
	   and Legs in Gauge Theories, Rheinsberg, Germany,
	   19-24 Apr 1998.}} \ubm
\author{M.~Passera and A.~Sirlin
\address{Department of Physics, New York University, \\
        4 Washington Place, New York, NY 10003, USA}}
\maketitle
\begin{abstract}
It is shown that conventional mass renormalization, when applied to
photonic or gluonic corrections to unstable particle propagators,
leads to non-convergent series in the resonance region. A solution of
this problem, based on the concepts of pole mass and width, is presented. 
In contrast with the $Z$ case, the conventional on-shell definition
of mass for $W$ bosons and unstable quarks contains an unbounded 
gauge dependence in next-to-leading
order. The on-shell and pole definitions of width are shown to
coincide if terms of $O(\Gamma^2)$ and higher are neglected, but not
otherwise.
\end{abstract}
\PACS{12.15.Lk,  11.10.Gh,  11.15.Bt,  14.70.Fm}
\vspace{-11cm}
\begin{flushright}
        \small
        NYU-TH/98-06-01\\
        June 1998
\end{flushright}  
\vspace{9cm}
%%%%%%%%%%%%%%%%%%%%%%%%%%%%%%%%%%%%%%%%%%%%%%%%%%%%%%%%%%%%%%%%%%%%%%%%%%%%%%%
\section{Introduction}
\pagestyle{myheadings}\markboth{~}{~}
Theoretical arguments advanced in 1991 led to the conclusion that,
in the $Z$ case, the on-shell mass
\be
        M^2 = M_0^2 + \mbox{Re} A(M^2),
\label{eq:M80}  
\ee
is gauge dependent in $O(g^4)$ and higher \cite{Si91a,Si91b}. If the
arguments are correct one should see the gauge dependence in the
analysis of the $Z$ resonant amplitude propagator or, equivalently, in
the study of the $Z$ line shape. This was, in fact, confirmed
\cite{PaSi96,Si-Ring}.
The situation in $O(g^4)$ is particularly simple to see. Calling 
$\widetilde{M}_{\smallz}$ the observed $Z$-mass, one finds 
\be
	\widetilde{M}_{\smallz}^2 = \mz^2 +\mz\Gamma_{\smallz}
		 \left[\mbox{Im}A'_b(\mz^2) - 
	\frac{9G_{\mu}m_b^2}{12 \pi \sqrt{2}} \right] + O(g^6),
\ee
where $\mz$ is the on-shell mass (Cf.~Eq.~(\ref{eq:M80})),
$A_b(\mz^2)$ is the bosonic contribution to the $Z$ self-energy,
the prime indicates differentiation with respect to $s$, and the term
proportional to $G_{\mu}m_b^2$ represents a very small violation to
the scaling behavior $\mbox{Im}A_f(s) \sim s$ ($A_f(s)$ is the
fermionic contribution to the self-energy). $\mbox{Im}A'_b(\mz^2)$ is
different from zero and $\xiw$-dependent when $\mz \geq 2\mw
\sqrt{\xiw}$ or $\xiw \leq 1/4\cos^2\theta_{\smallw}$. There is a second
class of contributions to  $\mbox{Im}A'_b(\mz^2)$ when 
$\mz \geq \mw(1+\sqrt{\xiw})$ or $\xiw \leq [(1/\cos\theta_{\smallw})-1]^2$.
One finds $|\mbox{Im}A'_b(\mz^2)|_{max}=1.6 \times 10^{-3}$ very near
the threshold of the second contribution. Thus, in $O(g^4)$ the gauge
dependence of $\mz$ 
is bounded and amounts to a maximum of
$2\mev$. Although very small, this is of the same magnitude as the
current experimental error. Instead, in $O(g^6)$ the gauge dependence
of $\mz$ is unbounded.

Another definition that plays an important role is based on the
complex-valued position of the pole \cite{var,Si91a,Si91b,PaSi96}:
\be
        \sov = M_0^2 + A(\sov), \quad \sov =m_2^2 -i m_2 \Gamma_2.
\label{eq:sov}
\ee
An important property is that $m_2$ and $\Gamma_2$ are gauge-independent.
In particular, 
\be
	m_1=\sqrt{m_2^2+\Gamma_2^2}
\label{eq:m1}
\ee
can be identified with the mass measured at LEP \cite{Si91a,Si91b}.

%%%%%%%%%%%%%%%%%%%%%%%%%%%%%%%%%%%%%%%%%%%%%%%%%%%%%%%%%%%%%%%%%%%%%%%%%%%%%%%
\bm\section{$W$ and Quark Propagators in the Resonance Region}\ubm
A very recent work has extended the analysis to $W$ and quark
propagators in the resonant region \cite{PaSi98}.

One finds that a new problem  emerges: in the treatment of the
photonic corrections, conventional mass-renormalization generates, in
next-to-leading order (NLO), a series in powers of $M\Gamma/(s-M^2)$,
which does not converge in the resonance region! Furthermore, it
diverges term-by-term at $s=M^2$. This problem is generally present
whenever the unstable particle is coupled to massless quanta. Aside
from the $W$, an interesting example is the QCD correction to a quark
propagator when the weak interactions are switched on, so that the quark
becomes unstable. In Ref.~\cite{PaSi98} a solution of this serious 
problem is presented in the framework of the complex pole formalism.

In order to illustrate the difficulties emerging in the resonance
region when conventional mass renormalization is employed, we consider
the contribution of the transverse part of the $W$ propagator in the
loop of Fig.~(1), which contains $l$ self-energy insertions.

%%%%%%%%% FIG.1
\begin{figure}[h]
  $$\epsfig{figure=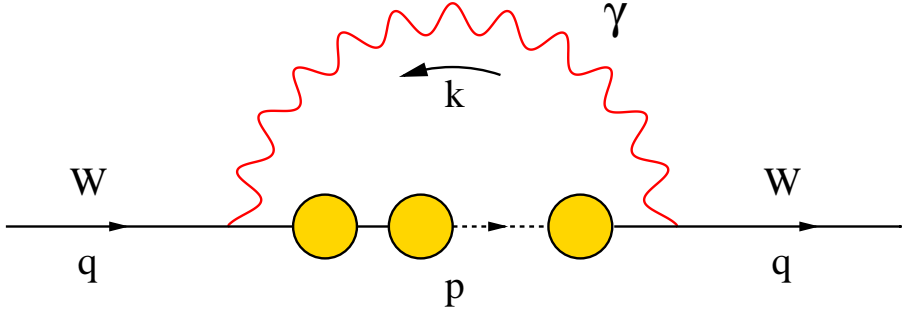,width=12cm}$$
  \caption{\sf A class of photonic corrections to the $W$ self-energy.
  The inner solid and dashed lines and blobs represent transverse
  $W$ propagators and self-energies.}
\end{figure}

\noindent Calling
\be
        \Pi^{\scriptscriptstyle (T)}_{\mu \nu}(q) = t_{\mu \nu}(q) A(s),
\ee
the transverse $W$ self-energy, where $s\equiv q^2$ and
$t_{\mu \nu}(q)=$ $g_{\mu \nu} - q_\mu q_\nu/ q^2$,
the contribution $A_{\smallw \gamma}^{(l)}(s)$ from
Fig.~(1) to $A(s)$ is given
by
\bea
        A_{\smallw \gamma}^{(l)}(s) &=& i e^2(\mu)
        \,\frac{t_{\mu \nu}(q)}{(n-1)}\, \mu^{4-n}
                                        \nonumber \\
        \times & & \!\!\!\!\!\!\!\!\!\!\!\!
        \int \frac{d^n \!k}{(2\pi)^n}
                {\cal{D}}
                ^{{\scriptscriptstyle (} \gamma {\scriptscriptstyle )}}
                _{\rho \beta} (k)
        {\cal{D}}^{\scriptscriptstyle (W,T)}_{\lambda \alpha}(p)
        {\cal{V}}^{\rho \lambda \nu} {\cal{V}}^{\beta \alpha \mu}
        \left[\frac{A^{{\scriptscriptstyle (} s {\scriptscriptstyle )}}(p^2)}
                {p^2 -M^2 +i \epsilon} \right]^l ,%\right\},
\label{eq:A-l-insertions}
\eea
where $p=q+k$ is the $W$ loop-momentum,
\be
        {\cal D}^{{\scriptscriptstyle (} \gamma {\scriptscriptstyle)}}
                _{\rho \beta} (k) = -\frac{i}{k^2}
        \left(g_{\rho \beta} + (\xig -1)
        \frac{k_\rho k_\beta}{k^2} \right),
\ee
\be
        {\cal D}^{\scriptscriptstyle (W,T)}_{\lambda \alpha}(p)=
        \frac{-i}{p^2 -M^2 +i\epsilon}
        \left(g_{\alpha \lambda} - \frac{p_\alpha p_\lambda}{p^2} \right),
\label{eq:Wprop-landau}
\ee
~
\be
        {\cal V}^{\beta \alpha \mu} = (2p-k)^\beta g^{\alpha \mu} +
        (2k-p)^\alpha g^{\beta \mu} -(k+p)^\mu g^{\beta \alpha},
\label{eq:vertex}
\ee
$\xig$ is the photon gauge parameter and
$A^{{\scriptscriptstyle (} s {\scriptscriptstyle )}}(p^2)$ stands for
the $W$ transverse self-energy with the conventional mass renormalization
subtraction:
\bea
        A^{{\scriptscriptstyle (} s {\scriptscriptstyle )}}(p^2) &=&
                        \mbox{Re}\left(A(p^2)-A(M^2)
                        \right) + i \mbox{Im}A(p^2)  \nonumber  \\
                &=& A(p^2) - A(M^2) + i\mbox{Im}A(M^2).
\eea
We note that each insertion of 
$A^{{\scriptscriptstyle (} s {\scriptscriptstyle )}}(p^2)$
is accompanied by an additional denominator
$[p^2-M^2 +i\epsilon]$. Thus, Eq.~(\ref{eq:A-l-insertions}) may be
regarded as the $l$th term in an expansion in powers of
$$
        \left[A(p^2) - A(M^2) + i\mbox{Im}A(M^2) \right]
        \left(p^2-M^2 +i\epsilon\right)^{-1}.
$$
As $A(p^2) - A(M^2) = O[g^2(p^2-M^2)]$ for $p^2 \approx M^2$, the 
contribution 
$[A(p^2) - A(M^2)]$ $(p^2-M^2 +i\epsilon)^{-1}$ is of $O(g^2)$ 
throughout the region of integration. However, as 
$i\mbox{Im}A(M^2) \approx -iM\Gamma$ is not subtracted, the combination
$i\mbox{Im}A(M^2)/(p^2-M^2 +i\epsilon)$ may lead to terms of $O(1)$
if the domain of integration $|p^2-M^2|$ \lsim $M \Gamma$ is important.
In fact, the contribution of $[i\mbox{Im}A(M^2)/(p^2-M^2 +i\epsilon)]^l$
to Eq.~(\ref{eq:A-l-insertions}) is, to leading order,
\be
        A_{\smallw \gamma}^{(l)}(s) =\frac{(-iM\Gamma)^l}{l!}
        \frac{d^l}{d(M^2)^l} A_{\smallw \gamma}^{(0)}(s) + \ldots,
\label{eq:A-l-insertions-Im}
\ee
where $A_{\smallw \gamma}^{(0)}(s)$ represents the diagram with no 
self-energy insertions and the dots indicate additional contributions
not relevant to our argument.

In the resonance region $|s-M^2|$ \lsim $M \Gamma$ the 
zeroth order propagator is inversely 
proportional to $(s-M^2+iM\Gamma) =O(g^2)$.
In NLO, contributions of $O[\alpha(s-M^2),\alpha M\Gamma]$ are therefore
retained but those of $O[\alpha(s-M^2)^2]$ are neglected.
Explicit evaluation of 
$A_{\smallw \gamma}^{(0)}(s)$ in NLO leads to 
\be
        A_{\smallw \gamma}^{(0)}(s) = \frac{\alpha}{2\pi}
        \left[(\xig-3)(s-M^2)\LM + \ldots\right].
\label{eq:A-0-insertions-partial}
\ee
Inserting Eq.~(\ref{eq:A-0-insertions-partial}) into 
Eq.~(\ref{eq:A-l-insertions-Im}) we obtain 
\bea
	A_{\smallw \gamma}^{(1)}(s) &=& \frac{\alpha}{2\pi}(\xig -3)
		\left( iM \Gamma \right) \left[ \LM + \fr{s}{M^2}
		\right] + \ldots,
			          \nonumber \\	
        A_{\smallw \gamma}^{(l)}(s) &=&\frac{\alpha}{2\pi}(\xig -3)
        \frac{(s-M^2)}{l(l-1)} 
        \left(\frac{-iM\Gamma\phantom{^2}}{s-M^2} \right)^{\!l} + \ldots, 
        \quad   (l \geq 2).
\label{eq:A-l-insertions-Im-again}
\eea
As in the resonance region all these terms contribute in NLO,
conventional mass renormalization leads in NLO to a series in powers
of $M\Gamma/(s-M^2)$, which does not converge in the resonance region.
Rather than generating contributions of higher order in $g^2$,
each successive self-energy insertion gives rise to a factor  
$-iM\Gamma/(s-M^2)$, which is nominally of $O(1)$ in the resonance region 
and furthermore diverges at $s=M^2$!

One possibility would be to resum the series 
$\sum^{\infty}_{l=0}A_{\smallw \gamma}^{(l)}(s)$ with 
$ A_{\smallw \gamma}^{(l)}(s)$ given by Eq.~(\ref{eq:A-l-insertions-Im}).
This would lead to 
\be
        \sum^{\infty}_{l=0} A_{\smallw \gamma}^{(l)}(s, M^2) = 
        A_{\smallw \gamma}^{(0)}(s, M^2-iM\Gamma) + \ldots,
\ee
or
\be
 	\sum^{\infty}_{l=0} A_{\smallw \gamma}^{(l)}(s)=\frac{\alpha}{2\pi}
        \left[(\xig-3)(s-M^2+iM\Gamma)\ln
        \left(\frac{M^2-iM\Gamma -s}{M^2-iM\Gamma}\right) + \ldots\right].
\label{eq:A-0-insertions-partial-again}
\ee
Even if one accepts these resummations rather than the usual term by
term expansions, the theoretical situation in the conventional
formalism is very unsatisfactory. 
In fact, in the conventional formalism, the $W$ propagator is inversely 
proportional to 
\be
        {\cal D}^{-1}(s) = s-M^2 +i M\Gamma 
		-\left( A(s)-A(M^2) \right) 
		-i M\Gamma \,\mbox{Re} A^\prime(M^2),
\label{eq:inverse_prop}
\ee
where $\Gamma$ is the radiatively corrected width and we have employed
its conventional expression
\be
	M \Gamma = -\mbox{Im}A(M^2)/[1-\mbox{Re}A'(M^2)].
\label{eq:usualwidth}
\ee 
The contribution of the 
$(s-M^2+iM\Gamma)\ln[(M^2-iM\Gamma -s)/(M^2-iM\Gamma)]$ term to 
${\cal D}^{-1}(s)$ is 
$$
        -\frac{\alpha}{2\pi} (\xig-3) \left[(s-M^2+iM\Gamma)\ln
        \left(\frac{M^2-iM\Gamma -s}{M^2-iM\Gamma}\right)
        +iM\Gamma \left(1+i\frac{\pi}{2} \right) \right]
$$
and we note that the last term is a gauge-dependent contribution not 
proportional to the zeroth order term $s-M^2+iM\Gamma$. 
As a consequence, in NLO the pole position is shifted to 
${\widetilde{M}}^2 -i\widetilde{M} \widetilde{\Gamma}$, where
\bea
	{\widetilde{M}}^2 &=& M^2[1-(\alpha/4)(\xig-3)(\Gamma/M)],
\label{eq:Mtilde}				\\
	\widetilde{\Gamma} &=& \Gamma [1-(\alpha/2\pi)(\xig-3)].
\label{eq:Gammatilde}	
\eea
As the pole position is gauge-invariant, so must be $\widetilde{M}$ and
$\widetilde{\Gamma}$. Furthermore, in terms of $\widetilde{M}$ and
$\widetilde{\Gamma}$, ${\cal D}^{-1}(s)$ retains the Breit--Wigner
structure. Thus, in a resonance experiment $\widetilde{M}$ and
$\widetilde{\Gamma}$ would be identified with the mass and width of $W$.

The relation $\widetilde{\Gamma} = \Gamma [1-(\alpha/2\pi)(\xig-3)]$
leads to a contradiction: the measured, gauge-independent, width 
$\widetilde{\Gamma}$ would differ from the theoretical value
$\Gamma$ by a gauge-dependent quantity in NLO! This contradicts the
premise of the conventional formalism that $\Gamma$, defined in
Eq.~(\ref{eq:usualwidth}), is the radiatively corrected width and is,
furthermore, gauge-independent. We can anticipate that the root of
this clash between the resummed expression and the conventional
definition of width is that the latter is only an approximation.
In particular, it is not sufficiently accurate when non-analytic
contributions are considered.

A good and consistent formalism may circumvent awkward resummations 
of non-convergent series and should certainly avoid the above
discussed contradictions.  
To achieve this, we return to the transverse dressed $W$ propagator,
inversely proportional to $p^2-M_0^2 -A(p^2)$. In the conventional
mass renormalization one eliminates $M^2_0$ by means of the expression
$M^2_0=M^2 -\mbox{Re}A(M^2)$ (Cf. Eq.~(\ref{eq:M80})). 
An alternative possibility is to
eliminate $M^2_0$ by means of $M^2_0=\sov -A(\sov)$
(Cf. Eq.~(\ref{eq:sov})).
The dressed propagator in the loop integral is inversely proportional
to $p^2 -\sov -[A(p^2)-A(\sov)]$. Its expansion about $p^2 -\sov$
generates in Fig.~(1) a series in powers of 
$[A(p^2)-A(\sov)]/(p^2-\sov)$.
As $A(p^2)-A(\sov)= O[g^2(p^2-\sov)]$ when the loop momentum is in the
resonance region, $[A(p^2)-A(\sov)]/(p^2 -\sov)$ is $O(g^2)$ 
throughout the domain of
integration. Thus, each successive self-energy insertion leads now to
terms of higher order in $g^2$ without awkward non-convergent
contributions.
In this modified strategy, the zeroth order propagator
in Eq.~(\ref{eq:Wprop-landau}) is replaced by 
\be
        {\cal D}^{\scriptscriptstyle (W,T)}_{\alpha \lambda }(p)=
        \frac{-i}{p^2 -\sov}
        \left(g_{\alpha \lambda} - \frac{p_\alpha p_\lambda}{p^2} \right).
\label{eq:Wprop-landau-sov}
\ee
The poles in the $k^0$ complex plane remain in the same quadrants as
in Feynman's prescription and Feynman's contour integration or 
Wick's rotation can be carried out.
$A_{\smallw \gamma}^{(0)}(s)$, Fig.~(1) without loop insertions, 
now leads directly to 
\be
        A_{\smallw \gamma}^{(0)}(s) = \frac{\alpha}{2\pi}
        \left[(\xig-3)(s-\sov)\LS + \ldots\right],	
\label{eq:A-0-insertions-partial-new}
\ee
which has the same structure as the expression we obtained in the
conventional approach after resumming a non-convergent series.
$A_{\smallw \gamma}^{(l)}(s)$ ($l\geq 1$), the contributions 
with $l$ insertions in Fig.~(1), are now of $O(\alpha g^{2l})$, 
the normal situation in perturbative expansions.
The $W$ propagator in the modified formalism is inversely
proportional to $s-\sov- [A(s)-A(\sov)]$. As $A_{\smallw
\gamma}^{(0)}(s)$ is now proportional to $s-\sov$, the pole position
is not displaced, the gauge-dependent contributions factorize as
desired, and the above discussed pitfalls are avoided. 
$A_{\smallw \gamma}^{(l)}(s)$ leads now to
contributions to $[A(s)-A(\sov)]$ of order  
$O[(s-\sov)\alpha g^{2l}] = O[\alpha g^{2(l+1)}]$ in the resonance
region and can therefore be neglected in NLO for $l\geq 1$.
We note that the $\ln [(\sov-s)/\sov]$ term in 
Eq.~(\ref{eq:A-0-insertions-partial-new}) cancels for $\xig=3$, the gauge
introduced by Fried and Yennie in Lamb-shift calculations 
\cite{Fried-Yennie}. 

The remaining contributions to $A(s)$ from the photonic diagrams,
including those from the longitudinal part of the $W$ propagator in 
Fig.~(1), and from the diagrams involving the unphysical scalars
$\phi$ and the ghost $C_\gamma$, have no singularities at $s=M^2$
and can therefore be studied with conventional methods. 
In particular, in the evaluation of $A(s)-A(\sov)$ in NLO it is
sufficient to retain their one-loop contributions. In these diagrams
the propagators are proportional to $(p^2-M^2\xiw)^{-1}$ rather than
$(p^2-M^2)^{-1}$. As a consequence, they lead to logarithmic terms
proportional to 
$$
        (s-M^2)\left[ \frac{s-M^2\xiw}{M^2} \right] \LMxiw.
$$ 
(The occurrence of branch cuts starting at $s=M^2\xiw$ indicates the 
unphysical nature of these singularities.) In the resonance region, in
NLO, these terms can be replaced by 
$(s-M^2)(1-\xiw)\ln[(\xiw-1)/\xiw]$ \cite{PaSi98}.

Calling $A^{\gamma}(s)$ the overall contribution of the 
one-loop photonic diagrams to the transverse $W$ self-energy 
(Fig.~(2)), in the modified
formulation the relevant quantity in the correction to the $W$
propagator is $A^{\gamma}(s) - A^{\gamma}(\sov)$. 
The corresponding one-loop gluonic contribution to
the quark self-energy is depicted in Fig.~(3).
In general $R_\xi$ gauge, we find in NLO
\bea 
        \lefteqn{\Agamma(s) - \Agamma(\sov) = 
                \frac{\alpha(m_2)}{2\pi} (s-\sov)
                \left\{\delta 
                \left(\frac{\xiw}{2}-\frac{23}{6}\right)
                +\frac{34}{9} -2\LS
                                \right.}                \nonumber \\
        & & \!\!\!\!\!\!\!\!\!\!
        -\left(\xiw-1\right)\left[\fr{\xiw}{12} -
                \left(1-\fr{(\xiw-1)^2}{12}\right)
                \ln \!\left(\fr{\xiw-1}{\xiw}\right) \right]      
                -\left(\fr{11}{12} 
                -\fr{\xiw}{4} \right) \ln\xiw
							\nonumber \\
        & & \!\!\!\!\!\!\!\!\!\!
        \left. +\left(\xig-1\right)\left[
                \fr{\delta}{2}+\fr{1}{2}
                +\LS+\fr{(\xiw^2-1)}{4}\ln\!\left(\fr{\xiw-1}{\xiw}\right) 
                -\fr{\ln\xiw}{4}+\fr{\xiw}{4} \right] \right\},
							\nonumber \\
\label{eq:Agamma-Agamma}                                                       
\eea
where $\delta = (n-4)^{-1} + (\gamma_E -\ln 4\pi)/2$, we have
treated the logarithmic terms according to the previous discussion
and set $\mu=m_2$. The full one-loop expression for $\Agamma(s)$ in
general $R_\xi$ gauges  without using the NLO approximation is given in 
Ref.~\cite{PaSi98}. Of particular interest in Eq.~(\ref{eq:Agamma-Agamma})
is the log term 
$$
	\frac{\alpha(m_2)}{2\pi} \left(\xig-3\right)
	\left(s-\sov\right)\LS,
$$
which is independent of $\xiw$ but is proportional to $(\xig-3)$.

%%%%%%%%% FIG.2 
\begin{figure}[h]
  $$\epsfig{figure=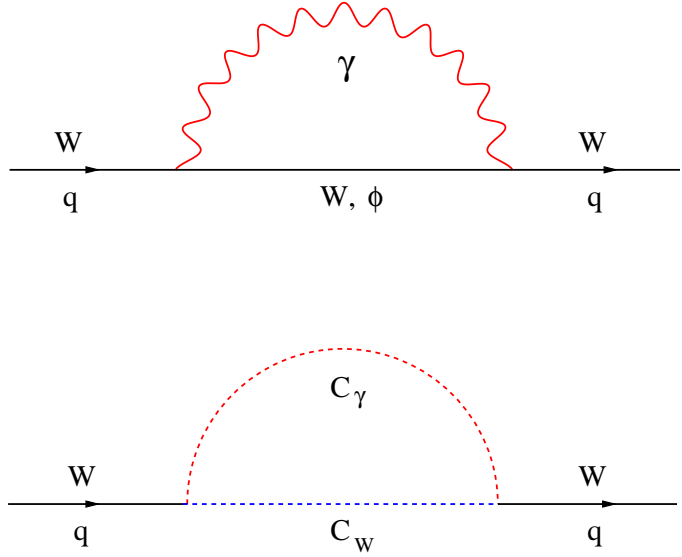,width=9cm}$$
  \caption{\sf One-loop photonic diagrams for the $W$ self-energy; 
  $\phi$ is the unphysical scalar, $C_\gamma$ and $C_W$ are ghosts.} 
  \label{figure:W-gamma-phi}
\end{figure}

Writing 
\be
        1-\fr{s}{\sov} = 1-\fr{s}{m_1^2}-i\fr{s}{m_1^2}
        \fr{\Gamma_2}{m_2} = \rho e^{i\theta},
\ee
we have
\be
        \rho (s) = \left[\left(1-\fr{s}{m_1^2}\right)^2 +
        \fr{s^2 \Gamma_2^2}{m_1^4 m_2^2} \right]^{1/2}, 
\label{eq:rho}
\ee
\be
        \rho \sin \theta (s)= -\fr{s \Gamma_2}{m_1^2 m_2}.
\label{eq:rhosintheta}
\ee
In Figs.~(4,5), 
the functions $\ln\rho(s)$ and $\theta(s)$ are plotted 
for $m_1=80.4\gev$ and $\Gamma_1=\Gamma_2 m_1/m_2=2\gev$
over a large range of $\sqrt{s}$ values. 
Figs.~(6,7)
compare these functions with the zero-width approximations 
over the resonance region. 
We note that the zero width approximation,
$$
        \mbox{Im}\left[\ln \!\left(\fr{M^2-s -i\epsilon}{M^2} 
        \right)\right]=  - \pi \theta\left(s-M^2\right),
$$
is not valid in the resonance region.
The logarithm $\ln(\xiw-1)$ in Eq.~(\ref{eq:Agamma-Agamma})
contains an imaginary contribution 
$-i\pi \theta(1-\xiw)$. This can be understood from the observation
that, for $\xiw<1$, a $W$ boson of mass $s=M^2$ has non-vanishing
phase space to ``decay'' into a photon and particles of mass $M^2\xiw$.

%%%%%%%%% Fig.3 -- QCD
\begin{figure}[h]
  $$\epsfig{figure=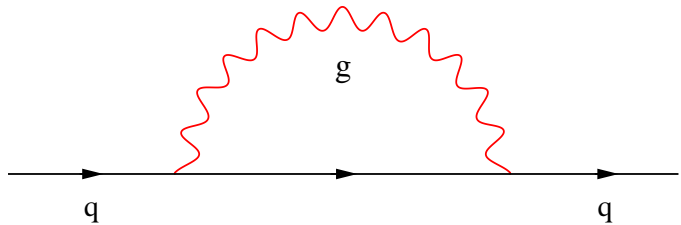,width=9cm}$$
  \caption{\sf One loop diagram for the quark self-energy in QCD.} 
  \label{figure:quark}
\end{figure}

%%%%%%%%%%%%%%%%%%%%%%%%%%%%%%%%%%%%%%%%%%%%%%%%%%%%%%%%%%%%%%%%%%%%%%%%%%%%%%%
\section{Gauge Dependence of the On-Shell Mass}

The difference between the pole mass $m_1$, defined in
Eq.~(\ref{eq:m1}), and the conventional on-shell mass $M$, 
defined in  Eq.~(\ref{eq:M80}), is
\be
        M^2 -m_1^2 = \mbox{Re} A(M^2) - \mbox{Re} A(\sov) -\Gamma_2^2.
\label{eq:deltaM}
\ee
The contribution of the $(s-\sov)\ln[(\sov-s)/\sov]$ term to the r.h.s. 
of Eq.~(\ref{eq:deltaM}) is
\bea
	& &\frac{\alpha(m_2)}{2\pi}\left(\xig -3\right)
	   \left[\left(M^2-m_2^2\right) 
		\mbox{Re}\ln \!\left(\frac{\sov-M^2}{\sov}\right)
		-m_2 \Gamma_2 \mbox{Im}\ln \! 
		\left(\frac{\sov-M^2}{\sov}\right)
	   \right]				\nonumber \\
	& \approx & \frac{\alpha(m_2)}{2\pi}\left(\xig -3 \right)
	   \left[\left(M^2-m_1^2\right)
		\mbox{Re}\ln \! \left(\frac{\sov-M^2}{\sov}\right)
		+m_2 \Gamma_2 \frac{\pi}{2}\right].
\eea
In $\mbox{Im}\ln[(\sov-M^2)/\sov]$ we have approximated 
$M^2 \approx m_1^2$ and used the fact that $\theta=-\pi/2$ for
$s= m_1^2$.
Thus, we have 
\be
        M^2 -m_1^2 = \fr{\alpha(m_2)}{4}(\xig-3)m_2\Gamma_2+ \ldots,
\label{eq:deltaMagain}
\ee
where the dots indicate additional contributions. We note that this last
equation corresponds to our previous result from the propagator, 
Eq.~(\ref{eq:Mtilde}), with the identification 
$\widetilde{M} \rightarrow m_1$.
In particular, Eq.~(\ref{eq:deltaMagain}) leads to 
$m_1-M=\alpha(m_2) \Gamma_2/4 \approx 4 \mev$ in the frequently 
employed 't Hooft--Feynman gauge $(\xi_i=1)$, and to $\approx 6\mev$
in the Landau gauge $(\xi_i=0)$.
The contribution to $M^2-m_1^2$ from the term proportional to 
$(s-\sov)(\xig-1)(\xiw^2-1)\ln(\xiw-1)$ (Cf.~Eq.~(\ref{eq:Agamma-Agamma})) 
is $(\alpha/8)(\xig-1)M\Gamma(\xiw^2-1)\,\theta(1-\xiw)$, which is 
unbounded in $\xig$ but restricted to $\xiw<1$.
In analogy with the $Z$ case, 
there are also bounded gauge-dependent contributions to $m_1-M$
arising from non-photonic diagrams in the restricted range
$\mw>\mz\sqrt{\xiz}+\mw\sqrt{\xiw}$ or
$\sqrt \xiz \leq \cos \theta_{\mbox{\footnotesize{w}}}[1-\sqrt \xiw]$,
and from the photonic corrections proportional to 
$(\xiw-1)\ln[(\xiw-1)/\xiw]$ (Cf. Eq.~(\ref{eq:Agamma-Agamma})).

The following observation is appropriate at this point. In calculating
the fundamental observable $\dlr$ \cite{Si80} (and its $\ms$ 
counterparts, $\Delta\hat{r}$ \cite{Si89-DFS91} and 
$\Delta\hat{r}_\smallw$ \cite{FaSi90}),   
the use of $M^2$ should produce a gauge dependence in $O(\alpha g^2)$
in the radiative corrections. How is this possible if $\dlr$
involves $A(s)\approx A(0)$? The point is that it also involves the
counterterm $\mbox{Re}A(M^2)$. If one employs the resummed expression 
$(\alpha/2\pi)(\xig-3)(s-M^2+iM\Gamma)\ln[(M^2-iM\Gamma-s)/(M^2-iM\Gamma)]$,
it gives a contribution $(\alpha/2\pi)(\xig-3)M\Gamma$ to 
$\mbox{Re}A(M^2)$. If one does not use the resummed expression, one
gets the same result from the graph of Fig.~(1) with one self-energy
insertion ($l=1$), provided one defines $\theta(0)=1/2$ in accordance
with the $i\epsilon$ prescription. One should eliminate such terms by
means of the replacement $M^2-(\alpha/4)(\xig-3)M\Gamma = m_1^2$ and
identify $m_1$ with the measured mass.

%%%%%%%%%%%%%%%%%%%%%%%%%%%%%%%%%%%%%%%%%%%%%%%%%%%%%%%%%%%%%%%%%%%%%%%%%%%%%%%
\bm \section{Overall Corrections to $W$ Propagators in the 
Resonance Region} \ubm

In contrast with the photonic corrections, the non-photonic
contributions $A_{np}(s)$ to $A(s)$ are analytic around $s=\sov$.   
In NLO we can therefore write 
\be
	A_{np}(s)-A_{np}(\sov) = (s-\sov)A_{np}'(m_2^2)+ \ldots,
\ee
where the dots indicate higher-order contributions. 

In the resonance region, and in NLO, the transverse $W$ propagator 
becomes
\be
	{\cal D}^{\scriptscriptstyle (W,T)}_{\alpha \beta}(q)=
        \frac{-i \left(g_{\alpha \beta} - q_\alpha q_\beta/ q^2\right)}
	     {\left(s-\sov\right)
		\left[1-A_{np}'(m_2^2) 
		-\frac{\alpha(m_2)}{2\pi}F(s,\sov,\xig,\xiw) \right]},
\label{eq:fullWprop-landau}
\ee
where $s=q^2$ and $F(s,\sov,\xig,\xiw)$ is the expression between
curly brackets in Eq.~(\ref{eq:Agamma-Agamma}). An alternative
expression, involving an $s-$dependent width, can be obtained by
splitting $A_{np}'$ into real and imaginary parts, and the latter into 
fermionic Im$A'_f$ and bosonic Im$A'_b$ contributions. 
Neglecting very small scaling violations, we have 
\be
	\mbox{Im} A'_f(m_2^2) \approx \mbox{Im} A_f(m_2^2)/m_2^2
			\approx -\Gamma_2/m_2
\ee
and 
\be
	{\cal D}^{\scriptscriptstyle (W,T)}_{\alpha \beta}(q)=
        \frac{-i \left(g_{\alpha \beta} - q_\alpha q_\beta/ q^2\right)}
	     	{\left(s-m_1^2 +is\frac{\Gamma_1}{m_1}\right)
	     	\left[1-\mbox{Re}A_{np}'(m_1^2)-i\mbox{Im}A_b'(m_1^2)
		-\frac{\alpha(m_1)}{2\pi}F \right] },
\label{eq:fullWprop-landau-again}
\ee
where $\Gamma_1/m_1=\Gamma_2/m_2$. $\mbox{Im}A_b'(m_1^2)$ is non-zero
and gauge-dependent in the subclass of gauges that satisfy 
$\sqrt \xiz \leq \cos \theta_{\mbox{\footnotesize{w}}}[1-\sqrt \xiw]$.
Otherwise $\mbox{Im}A_b'(m_1^2)$
vanishes. Although $m_1$ and $\Gamma_1$ are gauge-invariant, 
$\mbox{Re}A'_{np}(m_1^2)$, $\mbox{Im}A'_{np}(m_1^2)$
 and $F$ are gauge-dependent. In physical
amplitudes, such gauge-dependent terms cancel against contributions
from vertex and box diagrams.
The crucial point is that the gauge-dependent contributions in 
Eq.~(\ref{eq:fullWprop-landau-again}) factorize so that such
cancelations can take place and the position of the complex pole is
not displaced.

%%%%%%%%%%%%%%%%%%%%%%%%%%%%%%%%%%%%%%%%%%%%%%%%%%%%%%%%%%%%%%%%%%%%%%%%%%%%%%%
\bm \section{Comparison of the $W$ Width in the Conventional
and Modified Formulations} \ubm

Calling $A_0(s, M^2_0)$ the transverse self-energy evaluated in terms
of the bare mass $M_0$, and $A(s,M^2)$ and $\Aov(s,\sov)$ the
expressions obtained by substituting
$M_0^2=M^2-\mbox{Re}A(M^2,M^2)$ and $M_0^2=\sov -\Aov(\sov,\sov)$,
respectively, we have 
\be
	A_0(s, M^2_0) = A(s,M^2) = \Aov(s,\sov).
\label{eq:AAA}
\ee
In the conventional approach the $W$ width is given by
Eq.~(\ref{eq:usualwidth}) or, equivalently, 
\be
	M\Gamma = -\mbox{Im} A(M^2,M^2) + M \Gamma \,\mbox{Re}A'(M^2,M^2),
\label{eq:usualwidth-again}
\ee
where the prime means differentiation with respect to the first argument.
Instead, in the modified formulation, the width is defined by 
\be
	m_2 \Gamma_2 = -\mbox{Im} \Aov (\sov,\sov),
\label{eq:newwidth}
\ee
which follows from Eq.~(\ref{eq:sov}). If we combine 
Eq.~(\ref{eq:newwidth})
with Eq.~(\ref{eq:AAA}) and expand $\mbox{Im} A (\sov,M^2)$ about 
$\sov=M^2$, we find
\bea
	m_2 \Gamma_2 &=& -\mbox{Im} A (\sov,M^2) \nonumber  \\
		   &=& -\mbox{Im} A (M^2,M^2)
		       -\mbox{Im} \!\left[\left(
			\sov-M^2\right) A'(M^2,M^2) \right] + O(g^6).
						 \nonumber  \\
\label{eq:newwidth-again}
\eea
As $\sov -M^2 = m_2^2-M^2 -im_2\Gamma_2$ and $m_2^2-M^2=O(g^4)$, 
Eq.~(\ref{eq:newwidth-again}) becomes 
\be 
	m_2 \Gamma_2 = -\mbox{Im} A (M^2,M^2) +
		m_2\Gamma_2 \,\mbox{Re}A'(M^2,M^2)+ O(g^6).
\label{eq:newwidth-again2}
\ee
Comparing Eq.~(\ref{eq:usualwidth-again}) and Eq.~(\ref{eq:newwidth-again2}) 
we see that indeed 
\be
	\Gamma_2 = \Gamma + O(g^6).
\ee
Thus, the two calculations of the width coincide through $O(g^4)$,
i.e. in NLO. It is interesting to see how the two formulations treat
potential infrared divergences. 
In the conventional formulation, $\mbox{Re} A_{\gamma}'(M^2,M^2)$ 
is infrared divergent. This divergence is canceled by an 
infrared divergence in $\mbox{Im} A(M^2,M^2)$ arising from 
$A_{\smallw \gamma}^{(1)}(M^2,M^2)$, i.e. Fig.~(1)
with one self-energy insertion. 
In the modified expression $-\mbox{Im} \Aov (\sov,\sov)$ the two 
infrared divergent contributions are absent an one gets directly an
infrared convergent answer.

In high orders, if we insist in using the $(p^2-M^2+i\epsilon)^{-1}$
propagator and the conventional definition of the width, we are bound
to face severe infrared divergences due to the contributions of
Eq.~(\ref{eq:A-l-insertions-Im-again}) with $l\geq2$ which diverge as
powers in the limit $s \rightarrow M^2$. 
One could avoid this disaster by using the resummed expression 
$$
	\frac{\alpha}{2\pi}(\xig-3)(s-M^2+iM\Gamma)\
		\ln \! \left(\frac{M^2-s-iM\Gamma}{M^2-iM\Gamma} \right),
$$
but, as we saw earlier, this would give a contribution 
$(\alpha/2\pi)(\xig-3)M\Gamma$ to $M\Gamma$. In summary, the
conventional approach, based on the usual definition of width, is only
consistent if one neglects terms of $O(\Gamma^2)$ and higher. In the
modified formulation such problems don't arise. In particular, the
term $(\alpha/2\pi)(\xig-3)\ln[(\sov-s)/\sov]$ does not contribute to
the width.

%%%%%%%%%%%%%%%%%%%%%%%%%%%%%%%%%%%%%%%%%%%%%%%%%%%%%%%%%%%%%%%%%%%%%%%%%%%%%%%
\section{QCD Corrections to Quark Propagators in the Resonance Region}

In pure QCD quarks are stable particles, but  they become unstable
when weak interactions are switched on. As we anticipate similar 
problems to those in the $W$ case, we work from the outset in the
complex pole formulation.
Calling $\mov = m -i \Gamma/2$ the position of the complex pole,
$\Gamma$ arises from the weak interactions. If we treat $\Gamma$ to
lowest order, but otherwise neglect the remaining weak interactions
contributions to the self-energy, the dressed quark propagator can be
written
\be
	S'_F(\qsl) = \frac{i}{\qsl -\mov - 
		\left( \Sigma(\qsl) - \Sigma(\mov) \right)},  
\ee
where $\Sigma(\qsl)$ is the pure QCD contribution.
In NLO, in the resonance region, one finds
\be
	S'_F(\qsl) = \frac{i}{\left(\qsl -\mov \right)} 
		\left\{1-
		\frac{\alpha_s(m)}{3\pi} \left[2\left(\xi_g-3\right) 
			\ln \! \left(\frac{\mov^2-q^2}{\mov^2}\right)+
			2 \delta \xi_g \right]+\ldots \right\}^{-1},
\label{eq:S'_F}
\ee
where $\xi_g$ is the gluon gauge parameter and we have set $\mu=m$.
As in the $W-$propagator case, we see that the logarithm vanishes in
the Fried--Yennie gauge $\xi_g =3$.
The difference between $m$ and the on-shell mass
$M=m_0 +\mbox{Re}\Sigma(M)$ in leading order is 
\be 	
	M-m =\frac{\alpha_s(m)}{6}\Gamma \left(\xi_g-3\right),
\label{eq:deltaM-QCD}
\ee
which, in analogy with the $W$ case, is unbounded in NLO.
For the top quark, $m-M \approx 56\mev$ in the
Feynman gauge $(\xi_g=1)$, while in the Landau gauge
($\xi_g=0$) we have $m-M \approx 84\mev$.

%%%%%%%%%%%%%%%%%%%%%%%%%%%%%%%%%%%%%%%%%%%%%%%%%%%%%%%%%%%%%%%%%%%%%%%%%%%%%
\section{Conclusions}
The conclusions can be summarized in the following points.
i) Conventional mass renormalization, when applied to photonic and
gluonic diagrams, leads to a series in powers of $M\Gamma/(s-M^2)$
in NLO which does not converge in the resonance region.
ii) In principle, this problem can be circumvented by a resummation
procedure.
iii) Unfortunately, the resummed expression leads to an inconsistent
answer, when combined with the conventional definition of width.
This is not too surprising, as the traditional expression of width 
treats the unstable particle as an asymptotic state, which is clearly
only an approximation.
iv) An alternative treatment of the resonant propagator is
discussed, based on the complex-valued pole position 
$\sov= M_0^2+A(\sov)$.
The non-convergent series in the resonance region and the potential
infrared divergences in $\Gamma$ and $M$ are avoided by employing 
$(p^2-\sov)^{-1}$ rather than $(p^2-M^2)^{-1}$ in the Feynman integrals. 
The one-loop diagram leads now directly to the resummed expression of
the conventional approach, while the multi-loop expansion generates
terms which are genuinely of higher order. The non-analytic terms and
the gauge-dependent corrections cause no problem because they are
proportional to $s-\sov$ and therefore exactly factorize. 
v) The presence of $\sov$ in $\ln[(\sov-s)/\sov]$ removes the problem
of apparent infrared singularities.
vi) In contrast to the $Z$ case, the gauge dependence of the 
on-shell definition of mass for unstable $W$ bosons and quarks is
unbounded in NLO.
vii) It is shown that the conventional and modified definitions of
width coincide if terms of O($\Gamma^2$) and higher are neglected, but
not otherwise.

\newpage
%%%%%%%%% Fig.4 -- Log[rho[s]] 	-- W-wide range
$$\epsfig{figure=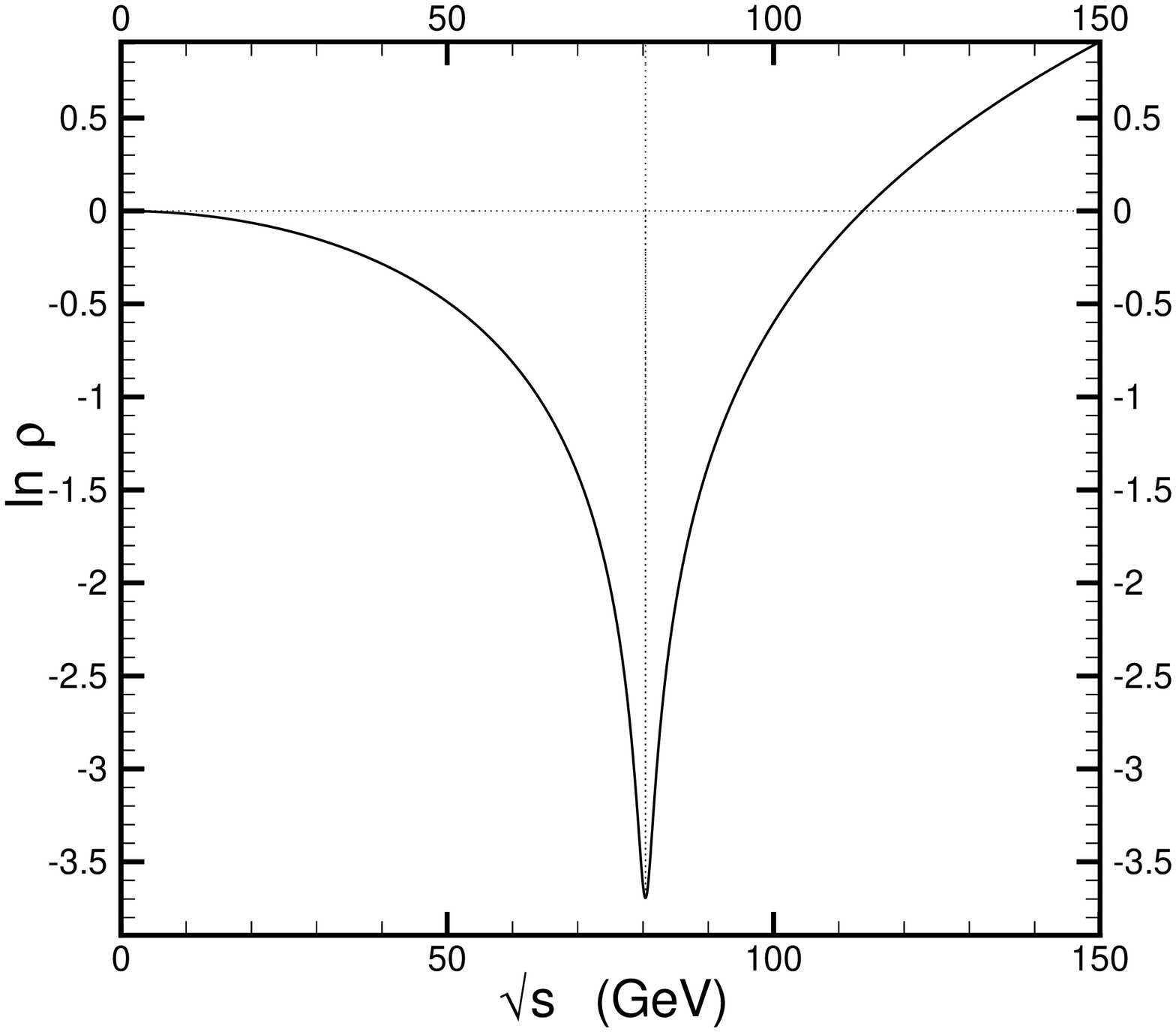,width=9cm}$$
\nobreak
Fig.4 {\sf The function $\ln\rho(s)$ over a large range of $\sqrt{s}$
values, for $m_1=80.4\gev$ and $\Gamma_1=2\gev$ 
(see Eq.~(\ref{eq:rho})). The minimum occurs at $\sqrt{s}=m_2$.}

%%%%%%%%% Fig.5 -- theta[s] 	-- W-wide range
$$\epsfig{figure=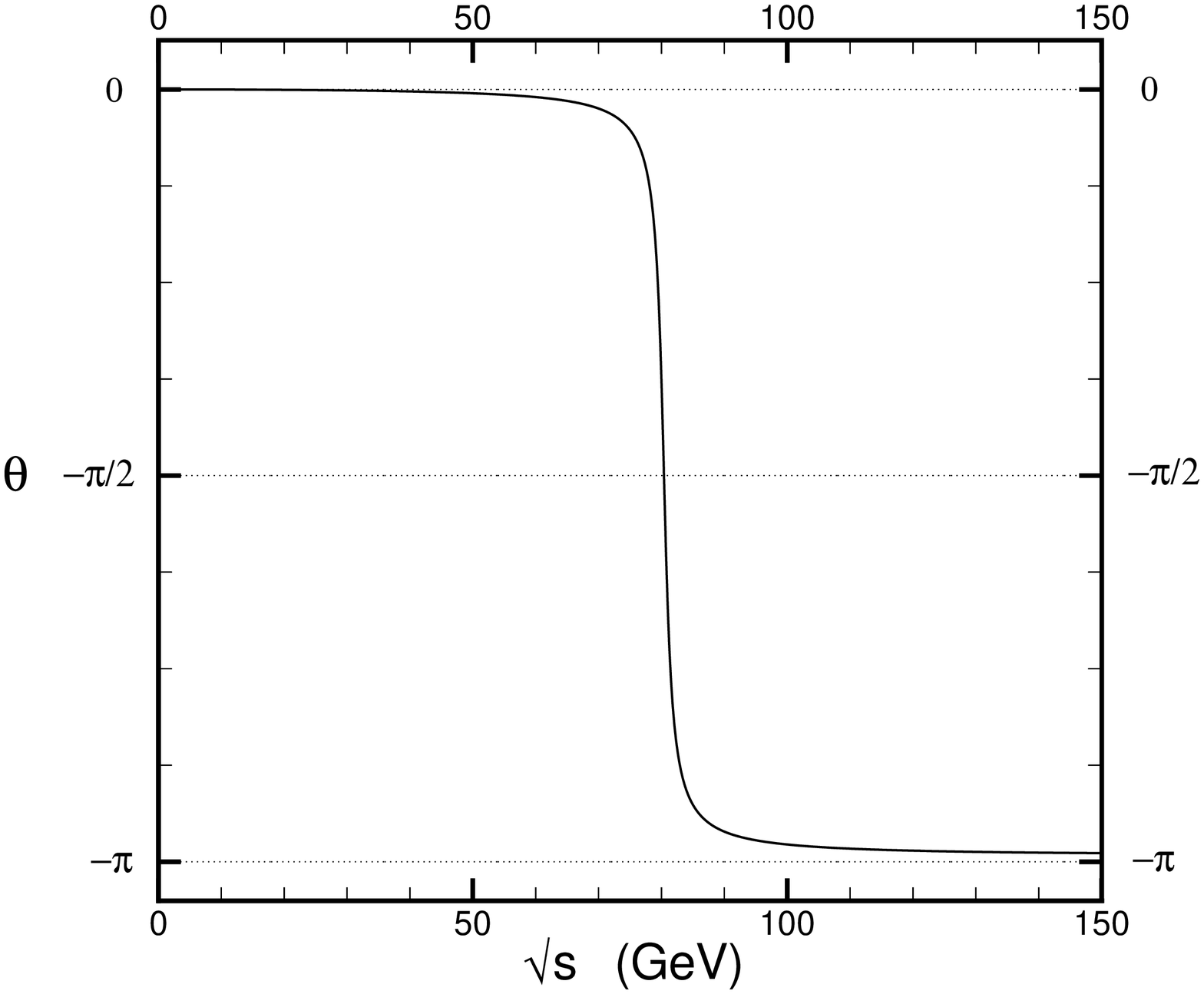,width=9cm}$$
\nobreak
Fig.5 {\sf The function $\theta(s)$ for $m_1=80.4\gev$ and
$\Gamma_1=2\gev$ (see Eq.~(\ref{eq:rhosintheta})). The value 
$-\pi/2$ is attained at $\sqrt{s}=m_1$.}

%%%%%%%%% Fig.6 -- Log[rho[s]]	-- W-res region 
$$\epsfig{figure=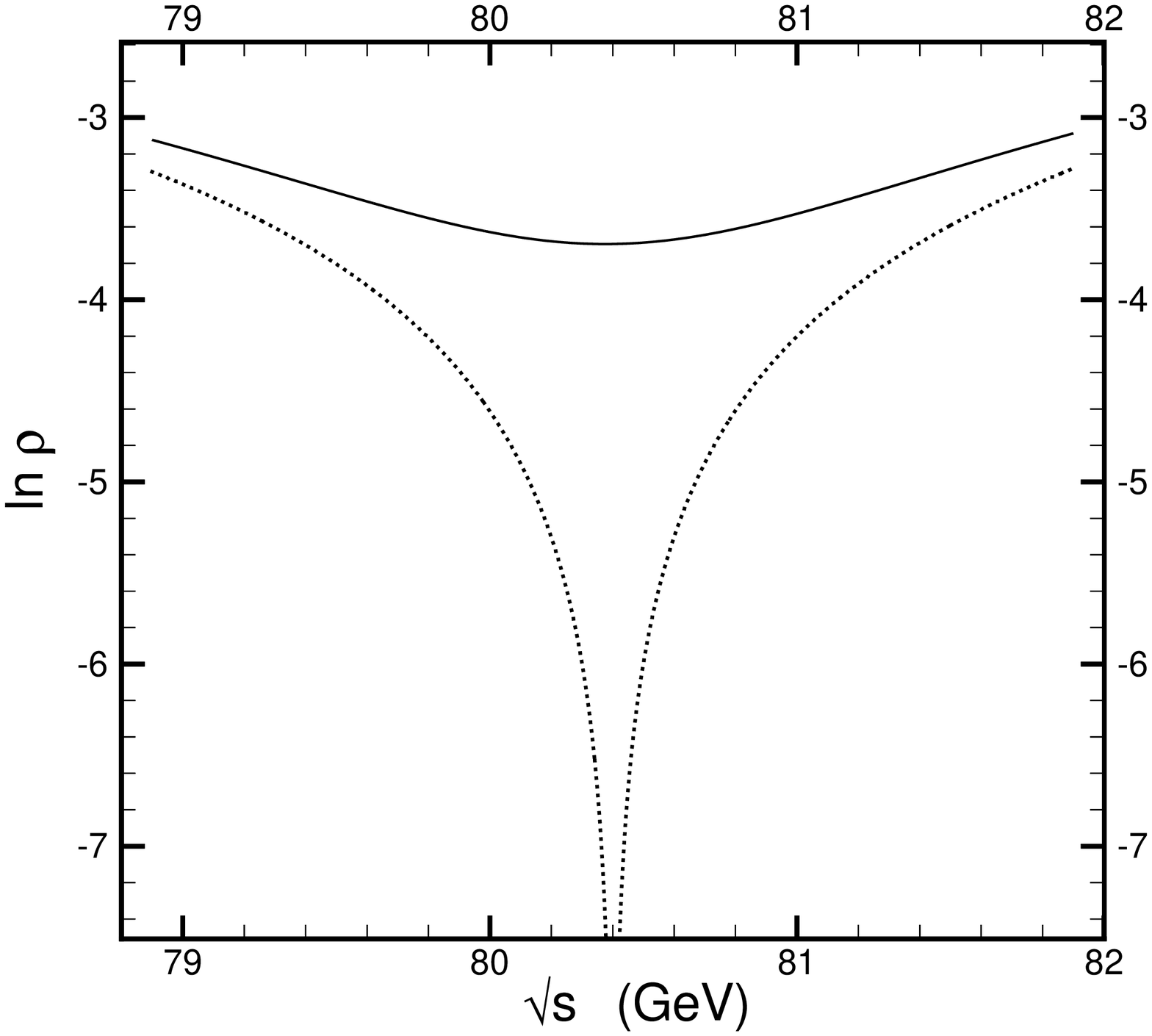,width=9cm}$$
\nobreak
Fig.6 {\sf Comparison of $\ln \rho(s)$ (solid line) with its zero-width 
approximation $\ln |1-s/m_1^2|$ (dotted line) over the resonance
region ($m_1=80.4\gev$, $\Gamma_1=2\gev$).}

%%%%%%%%% Fig.7 -- theta[s] 	-- W-res region 
$$\epsfig{figure=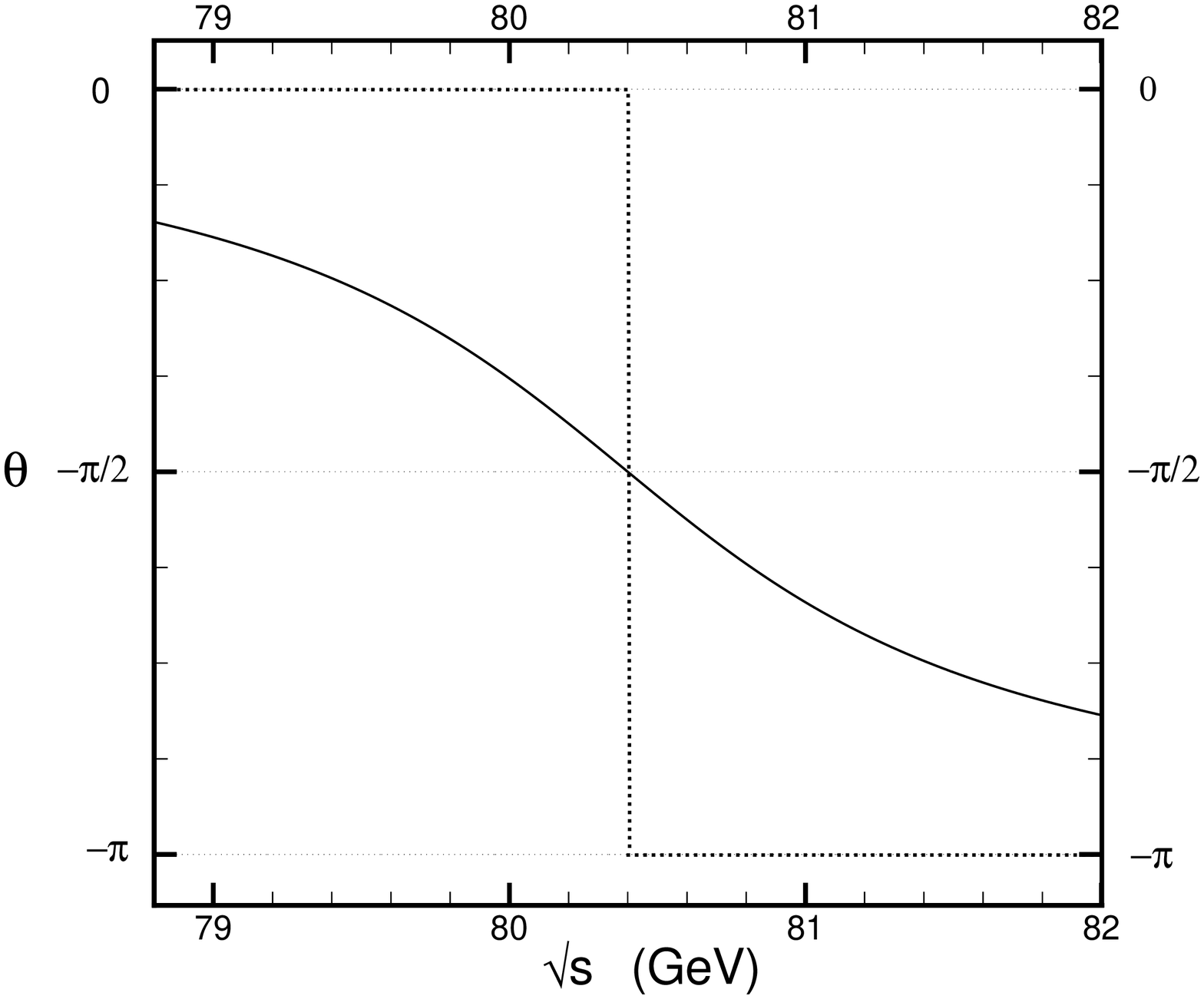,width=9cm}$$
\nobreak
Fig.7 {\sf Comparison of $\theta(s)$ (solid line) with the step function 
approximation (dotted line) over the resonance region
($m_1=80.4\gev$, $\Gamma_1=2\gev$).}

%%%%%%%%%%%%%%%%%%%%%%%%%%%%%%%%%%%%%%%%%%%%%%%%%%%%%%%%%%%%%%%%%%%%%%%%%%%%%

\end{document}